**Highly anisotropic electronic transport properties of monolayer and bilayer phosphorene from first principles**

Zhenghe Jin,[1] Jeffrey T. Mullen,[1] and Ki Wook Kim[1, 2, a)]

[1)]*Department of Electrical and Computer Engineering, North Carolina State University, Raleigh, NC 27695, USA*

[2)]*Department of Physics, North Carolina State University, Raleigh, North Carolina 27695, USA*

The intrinsic carrier transport dynamics in phosphorene is theoretically examined. Utilizing a density functional theory treatment, the low-field mobility and the saturation velocity are characterized for both electrons and holes in the monolayer and bilayer structures. The analysis clearly elucidates the crystal orientation dependence manifested through the anisotropic band structure and the carrier-phonon scattering rates. In the monolayer, the hole mobility in the armchair direction is estimated to be approximately five times larger than in the zigzag direction at room temperature (460 cm$^2$/Vs vs. 90 cm$^2$/Vs). The bilayer transport, on the other hand, exhibits a more modest anisotropy with substantially higher mobilities (1610 cm$^2$/Vs and 760 cm$^2$/Vs, respectively). The calculations on the conduction-band electrons indicate a comparable dependence while the characteristic values are generally smaller by about a factor of two. The variation in the saturation velocity is found to be less pronounced. With the anticipated superior performance and the diminished anisotropy, few-layer phosphorene offers a promising opportunity particularly in p-type applications.

---

[a)]Electronic mail: kwk@ncsu.edu

Two dimensional (2D) semiconducting materials have attracted much interest since the discovery of graphene due to their atomistically thin structure.[1] Though graphene provides extremely high mobility and electrical conductivity, its semi-metallic properties limit wide application in modern electronics.[2] Transition-metal dichalcogenides are, on the other hand, semiconductors with a graphene-like hexagonal structure in single or multiple layers.[3] Their band gaps, ranging approximately 1.0-2.5 eV, are normally at the K point in the Brillouin zone when direct.[3,4] Theoretical calculations predict intrinsic monolayer mobilities in the range of 25-410 $cm^2/Vs$ (electron) and 90-540 $cm^2/Vs$ (hole) for $MX_2$, where M=Mo,W and X=S,Se.[5–7] Experimentally reported values are substantially smaller with the highest around 250 $cm^2/Vs$ (p-type).[8]

Phosphorene, the single- or few-layer form of black phosphorous,[9] is another 2D semiconductor that has lately become a focus of investigation. High carrier mobility on the order of $10^5$ $cm^2/Vs$ has been observed in bulk phosphorous at low temperatures.[9,10] In the few-layer structures, on the other hand, the measurements as high as $\sim$220 $cm^2/Vs$ and 1000 $cm^2/Vs$ have been achieved for electrons and holes, respectively, under ambient conditions.[11,12] The reported values for monolayer (ML) and bilayer (BL) phosphorene are more modest at around 1 $cm^2/Vs$ and 80 $cm^2/Vs$ at low temperatures.[13] The crystal structure implies anisotropic characteristics for properties such as mechanical stability and electronic transport.[14,15] Further, the possibility of a tunable, direct band gap at the center of the Brillioun zone offers an interesting perspective for practical applications.[16] On the theoretical side, a detailed understanding of the carrier transport dynamics is yet to be achieved with the calculated ML electron mobilities ranging widely from 170 $cm^2/Vs$ to over 1000 $cm^2/Vs$ in the literature.[17–19] A careful analysis, particularly on the crucial carrier-phonon interactions, is evidently called for.

In this study, we investigate the intrinsic transport properties of ML and BL phosphorene from first principles. All of the key material characteristics (i.e., the electronic and phononic dispersion relations and the carrier-phonon interactions) are determined within the density functional theory (DFT) framework. The carrier transport is examined in the Boltzmann transport equation with a full-band Monte Carlo treatment. The results are compared with the available data in the literature and the relevant factors affecting the transport elucidated. The effective deformation potential constants are also extracted from the first-principles analysis.

In the calculation, a fully relaxed structure is first obtained by using the QUANTUM ESPRESSO package with ultrasoft pseudopotentials.[20] The minimum energy cutoff is set to 60 Ry in the plane-wave expansion along with the charge truncation of 360 Ry. The generalized gradient approximation is used for the exchange-correlation potential. The momentum space is sampled on a 12×12×1 grid for converged self-consistent calculations. The simulated cells are optimized until the atomic forces are less than 0.015 eV/Å. For ML phosphorene, the lattice constants are found to be $a = 4.50$ Å and $b = 3.23$ Å along the Γ-X and Γ-Y directions, respectively. The direct band gap of 0.91 eV is at the center of the Brillouin zone (Γ point) which is consistent with other theoretical reports in the literature.[11] For BL phosphorene, the interlayer separation between closest phosphorus atoms is 3.55Å while the in-plane lattice constant remains the same. The estimated BL band gap decreases dramatically to 0.63 eV. While DFT underestimates the band separation (see, for instance, a much larger gap of ∼ 2.0 eV predicted by the GW corrections),[21] this is expected to affect only the interband transitions between the conduction and the valence states. As such, the DFT band structure is deemed adequate for the present investigation. For convenience, Table I summarizes the effective masses extracted for both electrons and holes. An anisotropic behavior is evident in which the carriers are more massive in the zigzag direction (Γ-Y) than the armchair (Γ-X). The directional dependence appears more pronounced in the valence band. The full details of the electronic energy dispersion for both ML and BL phosphorene can be found in Ref. 22.

For the phonon dynamics and the carrier-phonon interaction, each phonon is treated as a perturbation of the self-consistent potential within the linear response [i.e., density functional perturbation theory (DFPT)].[20,23] The calculation of the potential change due to this perturbation gives the carrier-phonon interaction matrix element:[23,24]

$$g_{\mathbf{q},\mathbf{k}}^{(i,j)\nu} = \sqrt{\frac{\hbar}{2M\omega_{\nu,\mathbf{q}}}} \langle j, \mathbf{k}+\mathbf{q} | \Delta V_{\mathbf{q},\text{SCF}}^{\nu} | i, \mathbf{k} \rangle, \qquad (1)$$

where $|i, \mathbf{k}\rangle$ denotes the Bloch eigenstate with wave vector $\mathbf{k}$, band index $i$, and energy $E_{\mathbf{k}}^{i}$; $\Delta V_{\mathbf{q},\text{SCF}}^{\nu}$ is the change of the self-consistent Kohn-Sham potential due to the atomic displacement of a collective lattice vibration with wave vector $\mathbf{q}$ and frequency $\omega_{\nu,\mathbf{q}}$ (branch $\nu$); and $M$ is the atomic mass. The scattering rates are then obtained by the Fermi golden

rule:

$$\frac{1}{\tau_{\mathbf{k}}^i} = \frac{2\pi}{\hbar} \sum_{\mathbf{q},\nu,\pm} |g_{\mathbf{q},\mathbf{k}}^{i,\nu}|^2 (N_{\nu,\mathbf{q}} + \frac{1}{2} \pm \frac{1}{2}) \delta(E_{\mathbf{k}\mp\mathbf{q}}^i \pm \hbar\omega_{\nu,\mathbf{q}} - E_{\mathbf{k}}^i), \quad (2)$$

where $N_{\nu,\mathbf{q}}$ is the phonon occupation number following the Bose-Einstein statistics and $\pm$ represents phonon emission and absorption, respectively, In Eq. (2), only one band index ($i$) is adopted since the interband transitions are not considered hereafter. The obtained scattering rates are applied in the carrier transport studies along with a full-band treatment.

With a four-atom unit cell, ML phosphorene has 12 phonon branches.[15] Figure 1 details the scattering matrix elements obtained for a valence-band hole as a function of phonon wave vector $\mathbf{q}$ in the Brillouin zone. The two largest contributing modes, the longitudinal acoustic (LA) and $B_{3g}^1$ optical phonons,[25] are shown as an example interacting with the hole of energy $E_h = 50$ meV aligned along the high-symmetry axes in the momentum space, the armchair ($\Gamma$-X) and the zigzag ($\Gamma$-Y) [see (a,b) and (c,d), respectively]. The plots clearly illustrate the anisotropic dependence of the coupling strength not only on the phonon momentum $\mathbf{q}$ but also on the direction of the hole state $\mathbf{k}$. More specifically, the results show that the holes moving in the zigzag direction would interact more strongly with the phonons than those in the armchair direction. It is also indicated that the acoustic modes provide the dominant scattering mechanism compared to the optical phonons. The analysis for the conduction electron results in the analogous characteristics but with a substantially higher magnitude of interaction.[22]

Subsequent evaluation of hole scattering rates via phonon emission is shown in Fig. 2 as a function of energy. The anisotropic dependence on the crystallographic directions discussed above indeed leads to the disparate carrier relaxation dynamics. At very low energies, the holes with the momenta in the zigzag axis actually experience less interaction with both acoustic and optical phonons than the armchair counterparts. However, the coupling with the acoustic lattice oscillations rise rapidly with the hole energy above $10^{14}$ s$^{-1}$ for the former ($\Gamma$-Y, zigzag), while it remains relatively flat on the order of $10^{13}$ s$^{-1}$ for the latter ($\Gamma$-X, armchair). The optical modes play a comparatively minor role. Note also that only one onset (for emission) is visible here due to the absence of low-lying satellite valleys. The picture is similar for the phonon absorption scattering, where the zigzag case shows consistently higher rates (specifically, the contributions of the acoustic branches).

Figure 3 provides the velocity-field relation for ML phosphorene obtained from the Monte

Carlo simulations. The low-field mobilities for the holes are estimated to be 90 cm$^2$/Vs ($\mu_y$) and 460 cm$^2$/Vs ($\mu_x$) for the zigzag and armchair directions, respectively [Fig. 3(a)]. The ratio $\mu_x/\mu_y$ is approximately 5:1, suggesting a strongly anisotropic transport. The lower mobility along the zigzag direction is attributed to the larger effective mass and the stronger acoustic phonon scattering. The corresponding values in the conduction band are smaller by roughly a factor of two at around 40 cm$^2$/Vs ($\mu_y$) and 210 cm$^2$/Vs ($\mu_x$), while the anisotropy ratio turns out to be about the same [Fig. 3(b)]. Accordingly, it appears that the electrons are less efficient in the electrical conduction than the holes. This is despite the fact that the electron effective masses are comparable to or even smaller than those of the holes (see Table I). The stronger coupling with the phonons discussed earlier (particularly, in the zigzag direction) provides the explanation; viz., a sizable decrease in the relaxation time for the conduction-band electrons.[22] In contrast, an earlier theoretical investigation predicted the advantage of electrons over holes.[18] More specifically, their hole mobilities were somewhat lower than those of electrons. Interestingly, the results for the conduction-band transport are comparable in both calculations. As such, additional analyses may be necessary to clarify the point on the hole characteristics. In the experimental literature, the p-type samples tend to be reported with higher numbers.[11,26]

Similar studies have also been conducted for BL phosphorene with the outcome displayed in Fig. 4. The extracted mobilities are 1610 cm$^2$/Vs (armchair) and 760 cm$^2$/Vs (zigzag) for holes and 1020 cm$^2$/Vs (armchair) and 360 cm$^2$/Vs (zigzag) for electrons. Compared to the ML structure, the BL characteristics are clearly superior. In particular, the case with the lowest mobility (i.e., zigzag, conduction band) shows the relatively largest increase. Accordingly, the conduction becomes more isotropic between different crystallographic directions and carrier types. This is yet again a macroscopic consequence of the detailed carrier-phonon coupling/interaction dynamics.[22] The trend of improved performance in the thicker layers is consistent with the experimental observations in the p-type FET channels.[26] The smaller mobility values reported in the measurements can be attributed to additional scattering mechanisms, including ionized impurities and surface polar phonons, as well as a host of factors stemming from extrinsic conditions.

Along with the low-field transport, the velocity saturation behavior is examined as well. Inspection of the results in Figs. 3 and 4 shows that the drift velocity for the BL experiences a decrease after reaching the peak, while the behavior in the ML is rather monotonous.

Due to the sizable energy separation of the satellite valleys/peaks from the central minimum/maximum (over 200 meV) in both ML and BL structures,[22] it is unlikely that they would play a major role in the velocity drop at moderately high fields. One possible explanation for the disparate responses may be the large nonparabolicity observed in the BL. According to a discussion in the literature,[27] a strongly nonparabolic energy band can cause the negative differential mobility in the steady-state velocity-field characteristics. A more detailed investigation is needed for a definitive answer.

It is often convenient to approximate the ab initio carrier-phonon interaction with the deformation potential approximation. For ML and BL phosphorene, this is particularly relevant since only one valley/peak at the $\Gamma$ point plays the dominant role in the transport even for energetic carriers above 100 meV as mentioned above. Accordingly, the description can be limited to the intravalley transitions near the zone center. Then, the scattering rate by acoustic modes can be expressed in terms of the first-order deformation potential ($D_1$):

$$\frac{1}{\tau_{ac}^i} = \frac{m_d^i D_1^2 k_B T}{\hbar^3 \rho v_s^2}, \quad (3)$$

where $i$ denotes the band index as before ($c$,$v$), $m_d^i$ the corresponding density-of-states effective mass, $\rho$ the 2D mass density, and $v_s$ the sound velocity. Similarly, the interaction with the optical phonons is expressed with the zeroth order deformation potential ($D_0$) as:

$$\frac{1}{\tau_{op}^i} = \frac{m_d^i D_0^2}{2\hbar^2 \rho \omega_{op}}[N_{op} + (N_{op}+1)\Delta], \quad (4)$$

where $\Delta$ [$=\Theta(E_\mathbf{k}^i - \hbar\omega_{op})$] represents the Heaviside function for the emission process when the electron/hole energy $E_\mathbf{k}^i$ is greater than the phonon energy $\hbar\omega_{op}$. As usual, $N_{op}$ is the phonon occupation number following the Bose-Einstein statistics. In the calculation, $v_s$ is estimated from the LA phonon dispersion along the respective direction, while the geometric average of the $\Gamma$-X and $\Gamma$-Y effective masses is used for $m_d$. For simplicity, the contribution by all acoustic modes is attributed to one parameter $D_1$ without distinction. Likewise, the extracted $D_0$ accounts for the total scattering by the optical branches, for which a singe frequency of 28 meV is assumed at the zone center. The deformation potential values obtained for ML and BL are compiled in Table II.

In summary, the carrier-phonon interactions of ML and BL phosphorene are examined from first principles within the DFT and DFPT framework. Subsequent calculation of the scattering rates clearly illustrates the strong directional dependence that originates from the anisotropic coupling dynamics with the lattice. The estimated intrinsic transport properties are promising with the low-field mobility over 1000 cm$^2$/Vs and the saturation velocity around $10^7$ cm/s. It is also noted that a thicker layer may be more favorable than the ML case due to the higher mobilities and the reduced anisotropy.

This work was supported, in part, by SRC/NRI SWAN.

TABLE I. Calculated effective masses for ML and BL phosphorene in the armchair (Γ-X) and zigzag (Γ-Y) directions. The numerical values are given in units of electron rest mass $m_0$

|  | ML | | BL | |
|---|---|---|---|---|
|  | armchair | zigzag | armchair | zigzag |
| $m_\Gamma^c$ | 0.34 | 1.26 | 0.41 | 1.35 |
| $m_\Gamma^v$ | 0.30 | 3.02 | 0.33 | 4.76 |

TABLE II. Estimated deformation potential constants for hole-phonon and electron-phonon interactions in the highest valence band and the lowest conduction band, respectively, for ML and BL phosphorene. The acoustic (ac) sound velocities in the armchair and zigzag directions are $6.0 \times 10^5$ cm/s and $4.1 \times 10^5$ cm/s, respectively. For optical (op) phonons, a single energy of 28 meV at the zone center is used.

|  |  | phonon mode | deformation potential | |
|---|---|---|---|---|
|  |  |  | ML | BL |
| valence band | zigzag | ac (eV) | 4.3 | 1.5 |
|  |  | op (eV/cm) | $5.7 \times 10^8$ | $6.4 \times 10^8$ |
|  | armchair | ac (eV) | 7.1 | 4.3 |
|  |  | op (eV/cm) | $2.5 \times 10^9$ | $1.4 \times 10^9$ |
| conduction band | zigzag | ac (eV) | 15.0 | 2.9 |
|  |  | op (eV/cm) | $2.0 \times 10^9$ | $1.6 \times 10^9$ |
|  | armchair | ac (eV) | 7.9 | 5.5 |
|  |  | op (eV/cm) | $3.6 \times 10^9$ | $1.9 \times 10^9$ |

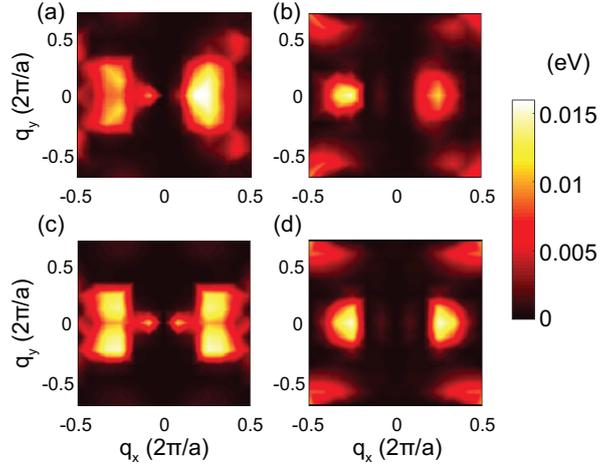

FIG. 1. Hole-phonon scattering matrix elements $|g^v_{\mathbf{q},\mathbf{k}}|$ in ML phosphorene (in units of eV) as a function of phonon wave vector $\mathbf{q}$ for the (a,c) LA and (b,d) $B^1_{3g}$ modes. The hole momentum $\mathbf{k}$ is assumed to be aligned along the armchair ($\Gamma$-X) and zigzag ($\Gamma$-Y) axes for (a,b) and (c,d), respectively, with an energy of 50 meV.

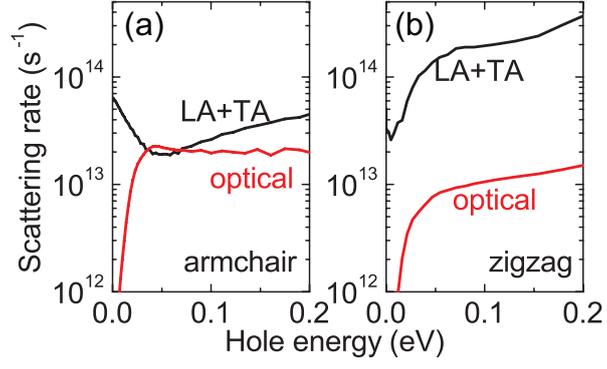

FIG. 2. Intrinsic scattering rates of holes in ML phosphorene via acoustic and optical phonon emission at room temperature. The hole momentum is assumed to be aligned along the (a) armchair ($\Gamma$-X) and (b) zigzag ($\Gamma$-Y) axes, respectively, as the energy varies.

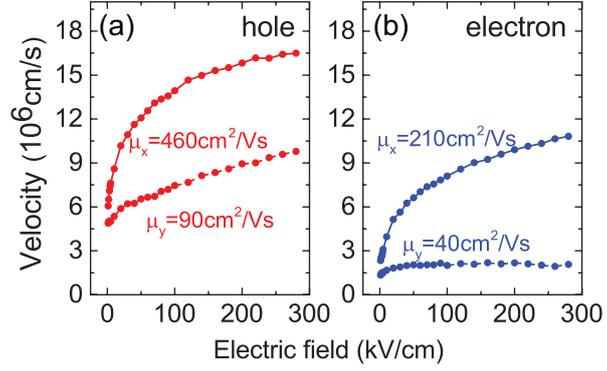

FIG. 3. (a) Hole and (b) electron drift velocities in ML phosphorene calculated as a function of the field applied in the armchair ($x$) and zigzag ($y$) directions.

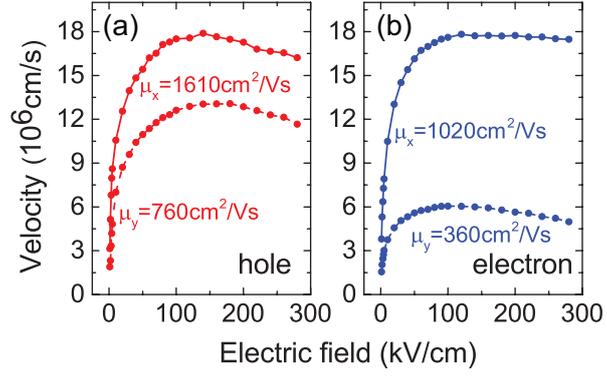

FIG. 4. (a) Hole and (b) electron drift velocities in BL phosphorene calculated as a function of the field applied in the armchair ($x$) and zigzag ($y$) directions.

# Supplementary: Highly anisotropic electronic transport properties of monolayer and bilayer phosphorene from first principles

Zhenghe Jin,[1] Jeffrey T. Mullen,[1] and Ki Wook Kim[1, 2, a]
[1] Department of Electrical and Computer Engineering, North Carolina State University, Raleigh, NC 27695, USA
[2] Department of Physics, North Carolina State University, Raleigh, North Carolina 27695, USA

## I. BAND STRUCTURE OF MONOLAYER AND BILAYER PHOSPHORENE

For an accurate description of the electronic and phononic states, the calculations are performed in the DFT framework as it is implemented in the QUANTUM-ESPRESSO package with ultrasoft pseudopotentials.[1] Figure S1 shows the electronic band structure of monolayer (ML) and bilayer (BL) phosphorene. The conduction- and valence-band extrema are located at the $\Gamma$ point in both cases with the estimated band gap of 0.91 eV and 0.63 eV, respectively. The precise values of the band gap are not crucial in the present study as discussed in the main paper. The energy differences between the central minimum (maximum) and the satellite valleys (peaks) are over 200 meV, essentially preventing the intervalley transitions in the low/moderate fields. The second band introduced by the BL is approximately 320 meV (490 meV) above (below) the lowest (highest) conduction (valence) band minimum (maximum) at the zone center.

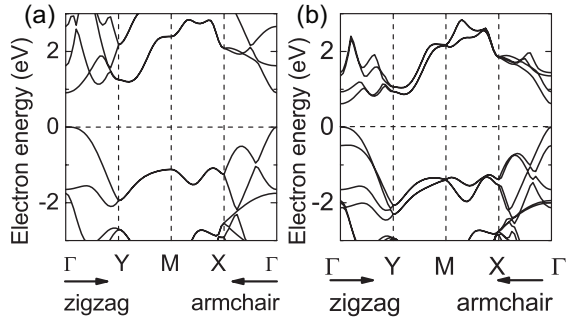

FIG. S1. Electronic band structures of ML and BL phosphorene along the symmetry directions in the Brillouin zone. The valence-band maximum at the $\Gamma$ point serves as the reference of energy scale.

## II. SCATTERING MATRIX FOR ELECTRONS IN MONOLAYER PHOSPHORENE

Along with the hole-phonon scattering matrix elements shown in the main paper (Fig. 1), the corresponding quantities are obtained for the conduction-band electrons as well (Fig. S2). A comparison clearly indicates that the electrons couple much more strongly with the phonons than the holes; note that the difference in the color coded scale in the two figures. The interaction with acoustic phonons remains distinctively anisotropic between the zigzag and armchair directions (as in Fig. 1). On the other hand, the anisotropy for the optical modes is substantially diminished.

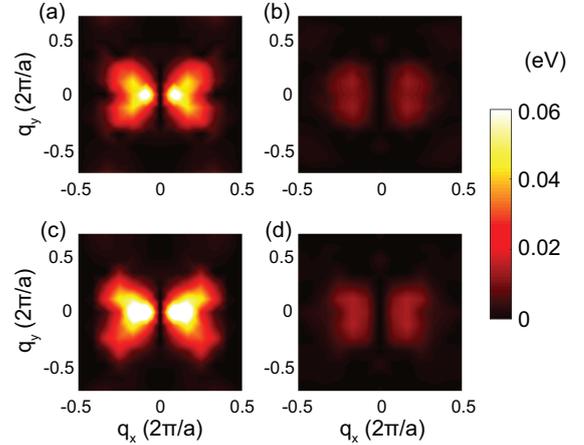

FIG. S2. Electron-phonon scattering matrix elements $|g^v_{\mathbf{q},\mathbf{k}}|$ in ML phosphorene (in units of eV) as a function of phonon wave vector $\mathbf{q}$ for the (a,c) LA and (b,d) $B^1_{3g}$ modes. The electron momentum $\mathbf{k}$ is assumed to be aligned along the armchair ($\Gamma$-X) and zigzag ($\Gamma$-Y) axes for (a,b) and (c,d), respectively, with an energy of 50 meV.

## III. INTRINSIC ELECTRON-PHONON SCATTERING RATES IN MONOLAYER PHOSPHORENE

Based on the matrix elements shown in Fig. S2, the intrinsic scattering rates for conduction-band electrons are calculated for ML phosphorene (Fig. S3). The results clearly illustrate that the electrons interact far more strongly with the phonons (particularly, the acoustic modes) than the holes in both the armchair and zigzag cases (see also Fig. 2 in the main paper). As mentioned earlier (Fig. S2), the directional dependence for the optical phonon scattering is apparently rather insignificant.

---

[a] Electronic mail: kwk@ncsu.edu

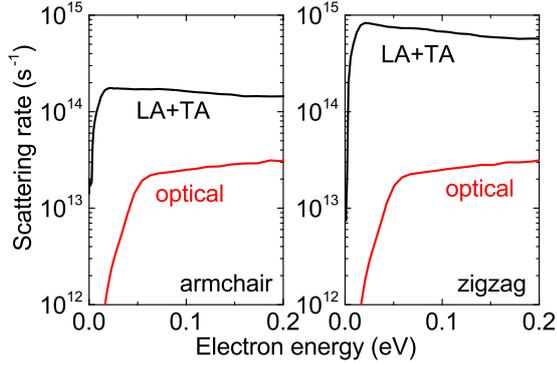

FIG. S3. Intrinsic scattering rates of conduction-band electrons in ML phosphorene via acoustic and optical phonon emission at room temperature. The electron momentum is assumed to be aligned along the armchair ($\Gamma$-X) and zigzag ($\Gamma$-Y) axes, respectively, as the energy varies.

## IV. SCATTERING RATES IN BILAYER PHOSPHORENE

For BL phosphorene, a similar picture can be constructed from first-principles calculations as described below. Figure S4 shows the hole scattering via phonon emission. A substantial reduction in the scattering rates is observed for all phonon branches compared to the ML structure (i.e., Fig. 2 in the main paper). Among the two transport directions, the zigzag case experiences a larger decrease in the interaction strength. Accordingly, the anisotropy tends to become more modest for holes in BL phosphorene, while the overall transport characteristics see a general improvement. Figure S5 provides the corresponding results for the conduction-band electrons. Here, the relative decline in the acoustic phonon scattering is even more pronounced for the zigzag structure.

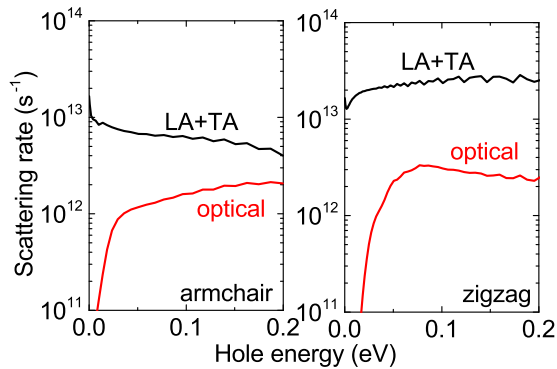

FIG. S4. Intrinsic scattering rates of valence-band holes in BL phosphorene via acoustic and optical phonon emission at room temperature. The hole momentum is assumed to be aligned along the armchair ($\Gamma$-X) and zigzag ($\Gamma$-Y) axes, respectively, as the energy varies. The small oscillations/fluctuations in the curves are numerical artifacts.

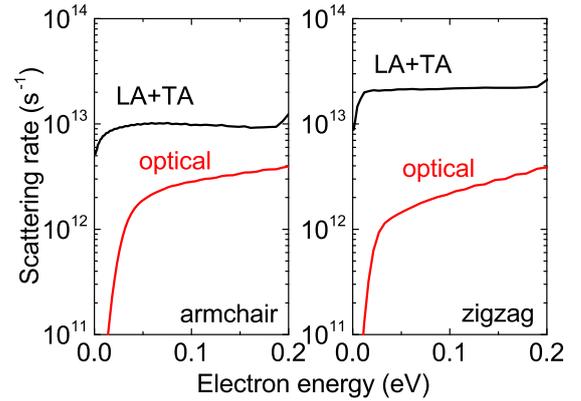

FIG. S5. Intrinsic scattering rates of conduction-band electrons in BL phosphorene via acoustic and optical phonon emission at room temperature. The electron momentum is assumed to be aligned along the armchair ($\Gamma$-X) and zigzag ($\Gamma$-Y) axes, respectively, as the energy varies.